\documentclass[10pt,conference]{IEEEtran}
\usepackage{makeidx}  
\usepackage{amssymb}
\usepackage{color}
\usepackage{graphicx}
\usepackage{subfigure}
\usepackage{url}
\usepackage{makecell}
\usepackage{cite}
\usepackage{listings}
\usepackage{paralist}
\let\subparagraph\paragraph
\usepackage{cite}
\usepackage{amsfonts, amsmath, amssymb}
\usepackage{epstopdf}
\usepackage{float}
\usepackage{enumerate}
\usepackage{color}
\usepackage{multirow}
\usepackage{verbatim}
\usepackage{stfloats}
\usepackage{xcolor}
\usepackage[utf8]{inputenc}
\usepackage{amsfonts}
\usepackage{eufrak}
\usepackage{eucal}
\usepackage{amsmath}
\usepackage{makeidx}
\usepackage{bm}
\makeatletter
\newif\if@restonecol
\makeatother

\usepackage[ruled,vlined, linesnumbered]{algorithm2e}
\makeatletter
\def\footnoterule{\kern-3\p@
  \hrule \@width 2in \kern 2.0\p@} 
\makeatother

\makeatletter
\def\BState{\State\hskip-\ALG@thistlm}
\makeatother

\definecolor{codegreen}{rgb}{0,0.6,0}
\definecolor{codegray}{rgb}{0.5,0.5,0.5}
\definecolor{codepurple}{rgb}{0.58,0,0.82}
\definecolor{backcolour}{rgb}{0.95,0.95,0.92}

\definecolor{dkgreen}{rgb}{0,0.6,0}
\definecolor{gray}{rgb}{0.5,0.5,0.5}
\definecolor{mauve}{rgb}{0.58,0,0.82}
\lstdefinestyle{mystyle}{
  backgroundcolor=\color{backcolour},   commentstyle=\color{codegreen},
  keywordstyle=\color{magenta},
  numberstyle=\tiny\color{codegray},
  stringstyle=\color{codepurple},
  basicstyle=\footnotesize,
  breakatwhitespace=false,
  breaklines=true,
  keepspaces=true,
  numbers=left,
  numbersep=5pt,
  showspaces=false,
  showstringspaces=false,
  showtabs=false,
  tabsize=2
}

\makeatletter
\newcommand{\xRightarrow}[2][]{\ext@arrow 0359\Rightarrowfill@{#1}{#2}}
\makeatother

\newcommand{\ccsl}{\textsc{Ccsl}}

\newcommand{\gt}[1]{\texttt{#1}}

\newcommand{\simu}{\textsc{Simulink}}
\newcommand{\staf}{\textsc{Stateflow}}

\newcommand{\ed}{\textsc{East-adl}}

\newcommand{\smt}{\textsc{Smt}}
\newcommand{\z}{\textsc{Z3}}
\newcommand{\zpy}{\textsc{Z3py}}
\newcommand{\py}{\textsc{Python}}
\newcommand{\sm}{\textsc{Sl}}
\newcommand{\sfl}{\textsc{Sf}}
\begin{document}

%

\title{SMT-based Probabilistic Analysis of Timing Constraints in Cyber-Physical Systems}
\author{
\IEEEauthorblockN{
Li Huang\IEEEauthorrefmark{1} and
Eun-Young Kang\IEEEauthorrefmark{1}\IEEEauthorrefmark{2}
 }
\IEEEauthorblockA{\IEEEauthorrefmark{1}School of Data \& Computer Science, Sun Yat-Sen University, Guangzhou, China \\
huangl223@mail2.sysu.edu.cn}
\IEEEauthorblockA{\IEEEauthorrefmark{2}PReCISE Research Centre,
University of Namur, Belgium\\
eykang@fundp.ac.be}
}
\maketitle

\begin{abstract}
Modeling and analysis of timing constraints is crucial in cyber-physical systems (CPS). \ed\ is an architectural language dedicated to safety-critical embedded system design. \simu/\staf\ (S/S) is a widely used industrial tool for modeling and analysis of embedded systems. In most cases, a bounded number of violations of timing constraints in systems would not lead to system failures when the results of the violations are negligible, called Weakly-Hard (WH).
We have previously defined a probabilistic extension of Clock Constraint Specification Language (\ccsl), called Pr\ccsl, for formal specification of \ed\ timing constraints in the context of WH.
In this paper, we propose an \smt-based approach for probabilistic analysis of \ed\ timing constraints in CPS  modeled in S/S: an automatic transformation from S/S models to the input language of \smt\ solver is provided; timing constraints specified in Pr\ccsl\ are encoded into \smt\ formulas and the probabilistic analysis of timing constraints is reduced to the validity checking of the resulting \smt\ encodings. Our approach is demonstrated a cooperative automotive system case study.
\end{abstract}

\section{Introduction}
Cyber-Physical Systems (CPS) are real-time embedded systems where the software controllers interact with physical environments. The continuous time behaviors of CPS often rely on complex dynamics as well as on stochastic behaviors.  Modeling and analysis of timing constraints is essential to ensure the correctness of CPS.
\ed \footnote{\scriptsize EAST-ADL. https://www.maenad.eu/public/EAST-ADL-Specification\_M2.1.9.1.pdf} is an architectural description language for safety-critical embedded systems design. The latest release of \ed\ has adopted the time model, which composes the basic timing constraints, i.e., repetition rates, end-to-end delays, and synchronization constraints.
\ed\ relies on external tools, e.g., \simu/\staf\footnote{\scriptsize Simulink and Stateflow. https://www.mathworks.com/products.html} (S/S), for system behaviors description.
\simu\ (\sm) is a block-diagram based formalism used to model continuous dynamics while \staf\ (\sfl) is used to specify control logic and state-based model behaviors of systems. Despite its strength in system modeling and simulation, S/S lacks of formal semantics to support rigorous verification of specifications.
To tackle this shortcoming, efforts have been devoted into formal analysis of S/S models by using formal methods,
e.g., model-checking, satisfiability modulo theory (\smt) solving.
However, the conventional formal analysis of real-time systems addresses worst case designs, typically used for hard deadlines in safety-critical systems. The ``Less-than-worst-case'' models are far less investigated.
In fact, in most cases, a bounded number of violations of timing constraints in
systems would not lead to system failures when the results of the
violations are negligible, called Weakly-Hard (WH)
\cite{Bernat2001Weakly}.
In this paper, we propose an \smt-based approach to support formal probabilistic analysis of \ed\ timing constraints in CPS modeled in S/S in the context of WH.

Clock Constraint Specification Language (\ccsl) is a formal language for specification of both logical and dense timing constraints.
We have previously defined a probabilistic extension of \ccsl, called Pr\ccsl\ \cite{setta18}, which states that the \emph{relations} (e.g., coincidence, causality and precedence) between events (e.g., input/output triggering, state changes) must hold with probability greater than or equal to a given probability threshold. Previous work is extended by including the supports of probabilistic analysis of timing constraints using \smt-based model checking:
\begin{inparaenum}
\item S/S models, which describe the behaviors of systems, are transformed into the input language of \smt\ solver;
\item \ed\ timing constraints with stochastic properties are specified in Pr\ccsl\ and encoded into \smt\ formulas;
\item The probabilistic analysis of timing constraints is reduced into validity checking of the resulting \smt\ encodings.
\end{inparaenum} Our approach is demonstrated on a cooperative automotive system case study.

\section{Methodology \& Experiment}
\begin{figure}[htbp]
  \centering
  \includegraphics[width=3.5in]{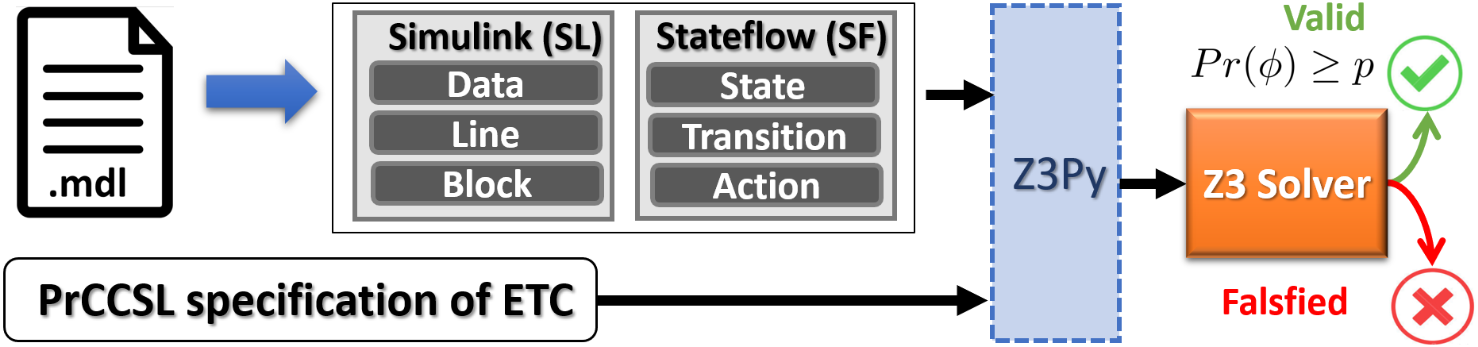}\\
  \caption{Overview of our approach}
  \label{fig:app}
  \end{figure}

The overview of our approach is shown in Fig. \ref{fig:app}.
In our approach, S/S models are stored in `.mdl' files, which contain textual descriptions of the compositions of the models. \z\ \smt\ solver\footnote{\scriptsize Z3 SMT Solver. https://github.com/Z3Prover/z3} is employed as our verification engine.
To translate the stochastic functions (e.g., random number generation) in \sm, we adapt \zpy\   (the \z\ API in \py) as encoding interface,
in which the add-on modules for description of probability distributions can be leveraged.
Given a system model in S/S and an \ed\ timing constraint $\phi$ (specified in Pr\ccsl),
the goal of our approach is to verify whether the probability of the constraint is greater than or equal to a probability threshold $p$, i.e., $Pr(\phi)\geq p$.  To achieved this, we perform the following steps:
\begin{inparaenum}
\item Extract necessary information (see Fig. \ref{fig:app}) of S/S from .mdl file and translate S/S into \zpy\ encodings based on the extracted information;
\item Encode Pr\ccsl\ specifications of \ed\ timing constraints (ETC) in \zpy\ and check the validity of the encodings using \z.
\end{inparaenum}

\begin{figure}[htbp]
  \centering
  \includegraphics[width=3.5in]{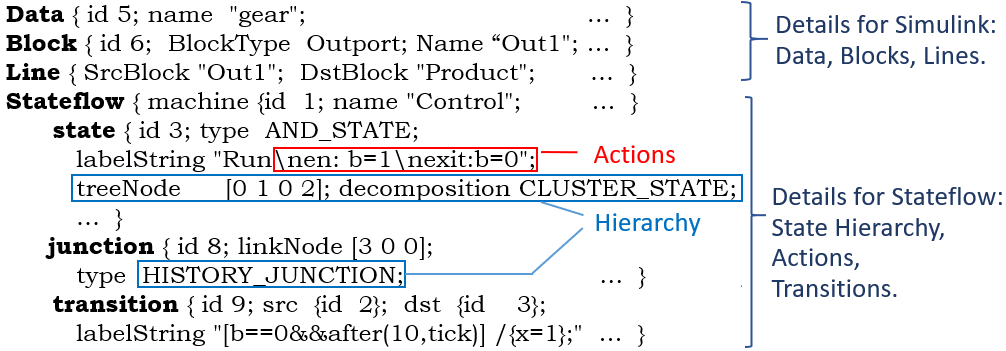}\\
  \caption{S/S information in .mdl file}
  \label{fig:mdl}
  \end{figure}

\vspace{0.05in}
\noindent\textbf{Translation of S/S into \zpy:} Fig. \ref{fig:mdl} shows an excerpt of S/S model in .mdl file, in which each object (e.g., block, data or state) has a unique identifier named $id$.
The data/variables in discrete-time S/S model are updated at $sample\ time$ steps, which are translated into vectors (i.e., bounded lists) of appropriate sorts (e.g., integer, real and boolean). The index of the vectors represent the number of time steps have proceeded during simulation.
For instance, an integer signal $a$ is mapped to an integer list, with $a[i]$ ($i$$\in$$\mathbb{N}$) representing the value of signal $a$ at $i^{th}$ step during simulation.
In \sm, lines are used for data transmission.  During simulation, the data of ports connected by the same line are identical, which is interpreted as the equivalence of the data in \zpy.
The blocks of linear math/logic functions in \sm\ are mapped to the same arithmetic/logical operations in \zpy\ straightforwardly.
\vspace{-0.2in}
\vspace{-0.15in}

\emph{States} in \sfl\ can be either active or inactive during simulation, declared as integer vectors whose elements are either 1 (active) or 0 (inactive) in \zpy.
The information of $state$ in .mdl file can be divided into three classes (see Fig. \ref{fig:mdl}):
\emph{Hierarchy} includes \emph{decomposition}, \emph{history junction} and the relation between \emph{superstates} and their \emph{substates} (indicated by \emph{treeNode}).
\emph{Transition} represents the passage of the system from one state to another when the condition (i.e., a boolean expression) on the transition is true.
\emph{State action} refers to the operations (e.g., assignments) executed when the state is active, entered or exited.
After the information of hierarchy, transitions and actions (HTA) is extracted from the .mdl file, the translation of \sfl\ becomes the interpretation of HTA of each state in \zpy, as presented in Algorithm \ref{alg:r2p}.
\begin{algorithm}[htbp]
    \small
    \caption{\small{Translation of \staf\ into \zpy}}
    \label{alg:r2p}
    \KwIn{Simulation bound $N$, states $s_0, ..., s_j, ..., s_n$ ($s_j$ is the state with $id$ $j$), information of hierarchy,  actions and transitions of each state;}
    \KwOut{$E$: the \zpy\ encodings of \staf\ chart;}
    $E$ $\leftarrow$ $\varnothing$\\
    \For{$j = 0; j< n; j ++$}
    {$s_j$ $\leftarrow$ IntegerVector($N$) \\
    \If{$s_j.substate \neq \varnothing$} {$E$ $\leftarrow$ $E$ $\cup$ Encode($s_j.hierarchy$)}
    \vspace{-0.05in}
     \If{$s_j.action \neq \varnothing$}{$E$ $\leftarrow$ $E$ $\cup$ Encode($s_j.action$)}
     \vspace{-0.05in}
     \If{$s_j.transition \neq \varnothing$}{$E$ $\leftarrow$ $E$ $\cup$ Encode($s_j.transition$)}
     \vspace{-0.1in}
    }
    return $E$
    \vspace{-0.05in}
\end{algorithm}
\setlength{\textfloatsep}{3pt}

\vspace{0.05in}
\noindent\textbf{Experiment:}
Our approach is demonstrated on a cooperative automotive system (CAS) \cite{sac18}, which includes distributed and coordinated sensing, control, and actuation over two vehicles running in the same lane.
We consider seven types of timing constraints of CAS system, i.e., {\small \gt{End-to-End}}, {\small \gt{Periodic}}, {\small \gt{Sporadic}}, {\small \gt{Execution}}, {\small \gt{Synchronization}}, {\small \gt{Comparison}} and {\small \gt{Exclusion}} constraints. The timing constraints are specified in Pr\ccsl, whose semantics is specified in the form of \smt\ formulas \cite{setta18} that can be expressed in \zpy\ naturally.
In our experiment, the simulation bound and the probability threshold are set to 3000 steps and 95\% respectively. 
The experiment results are listed in Table \ref{table_verification_result} and all timing constraints are established as valid.
\vspace{-0.1in}
\begin{table}[htbp]
 \scriptsize
  \centering
   \renewcommand\arraystretch{1.3}
  \caption{Verification Results of Timing Constraints in Z3}
    \begin{tabular}{|c|c|c|c|c|}
    \hline
    Timing Constraint  & Results & Time (Min) & Mem (Mb) & CPU (\%) \\
    \hline
     End-to-End & valid & 70.3 & 1710.75 & 23.81
         \\ \hline

     Periodic & valid & 12.7 & 2639.25 & 24.86
         \\ \hline

     Sporadic & valid & 202.4 & 1869.02 & 24.89
         \\ \hline
     Execution & valid & 12.7 & 2516.13 & 24.89
         \\ \hline
     Synchronization & valid & 63.6 & 2299.83 & 20.36
         \\ \hline

     Comparison & valid & 65.7 & 2005.06 & 23.63
         \\ \hline

     Exclusion & valid & 112.4 & 2569.08 & 20.47
    \\ \hline
    \end{tabular}%
  \label{table_verification_result}%
\end{table}%
\vspace{-0.1in}
\section{Conclusion and Future Work}
We present an \smt-based approach to perform probabilistic analysis of \ed\ timing constraints in CPS described in \simu/\staf. The practicality of our approach is demonstrated on a CAS case study.
As ongoing work, the application of our approach in larger-scale case studies will be investigated and an automatic translator from \simu/\staf\ (.mdl file) to \zpy\ will be developed.

\bibliographystyle{IEEEtran}
\bibliography{reference}

\end{document}